\documentclass[aps, prl, twocolumn, groupedaddress, floatfix]{revtex4-1}

\RequirePackage{color}
\RequirePackage[colorlinks, urlcolor=my-blue,linkcolor=my-red,citecolor=my-green]{hyperref}
\definecolor{my-blue}{rgb}{0.0,0.0,0.6}
\definecolor{my-red}{rgb}{0.5,0.0,0.0}
\definecolor{my-green}{rgb}{0.0,0.5,0.0}
\definecolor{nicos-red}{rgb}{0.75,0.0,0.0}
\definecolor{light-gray}{gray}{0.6}
\definecolor{really-light-gray}{gray}{0.8}
\definecolor{sussexg}{rgb}{0.0,0.5,0.5}
\definecolor{sussexp}{rgb}{0.5,0.0,0.5}

\usepackage{nicefrac, natbib, amsmath, amssymb, graphics,tikz,epstopdf,array,epstopdf,stackengine}

\newcommand{\be}{\begin{equation}}
\newcommand{\ee}{\end{equation}}

\usepackage[british]{babel}

\def\m1{\mathbf{1}}

\definecolor{darkgreen}{rgb}{0.0,0.5,0.0}
\definecolor{darkblue}{rgb}{0.0,0.0,0.3}
\definecolor{nicosred}{rgb}{0.65,0.1,0.1}
\definecolor{light-gray}{gray}{0.7}

\begin{document}

\title{Pairwise-like models for non-Markovian epidemics on networks}

Istvan Z. Kiss $^{1,\ast}$, Gergely R\"ost$^{2}$ \& Zsolt Vizi $^{2}$

\author{Istvan Z. Kiss}
\email[]{i.z.kiss@sussex.ac.uk}
\affiliation{School of Mathematical and Physical Sciences, Department of Mathematics, University of Sussex, Falmer, Brighton BN1 9QH, UK}

\author{Gergely R\"ost}
\email[]{rost@math.u-szeged.hu}
\affiliation{Bolyai Institute, University of Szeged, Aradi v\'ertan\'uk tere 1, Szeged 6720, Hungary}

\author{Zsolt Vizi}
\email[]{zsvizi@math.u-szeged.hu}
\affiliation{Bolyai Institute, University of Szeged, Aradi v\'ertan\'uk tere 1, Szeged 6720, Hungary}

\date{\today}

\begin{abstract}
%
In this letter, a generalization of pairwise models to non-Markovian epidemics on networks is presented. 
For the case of infectious periods of fixed length, the resulting pairwise model is a system of delay differential equations (DDE), which shows 
excellent agreement with results based on explicit stochastic simulations of non-Markovian epidemics on networks. Furthermore, 
we analytically compute a new $\mathcal{R}_0$-like threshold quantity and an implicit analytical relation between this and the final epidemic size. 
In addition we show that the pairwise model and the analytic calculations can be generalized in terms of integro-differential equations
to any distribution of the infectious period, and we illustrate this by presenting a closed form expression for the final epidemic size. 
By showing the rigorous mathematical link between non-Markovian network epidemics and pairwise DDEs, we provide
the framework for a deeper and more rigorous understanding of the impact of non-Markovian dynamics
with explicit results for final epidemic size and threshold quantities.\\

%
%

\end{abstract}


\maketitle


Networks have provided a step change in modelling complex systems ranging from 
man-made to natural ones \cite{newman2003structure,boccaletti2006complex,pastor2014epidemic}. 
The study of disease transmission on networks has particularly benefitted 
from this modelling paradigm by uncovering the role and impact of contact 
heterogeneity and clustering \cite{pastor2014epidemic}, to name just a few.
While networks provide a clear departure from classic compartmental models,
the role of mean-field models remains crucial. 
These offer a reliable way to obtaining analytical results 
and thus uncovering the interplay between network properties and the dynamic 
processes on networks. For example, the epidemic threshold \cite{SatorrasVespignani,liurostvas} and 
final epidemic size \cite{keeling1999effects} can be given in terms of explicit or implicit mathematical expressions 
which clearly illustrate how network and disease parameters combine.

Probably the most widely spread and well-known mean-field model for network epidemics is the degree-based mean-field (DBMF) model, also known as heterogeneous mean-field \cite{SatorrasVespignani,pastor2014epidemic}.
Similarly, pairwise models \cite{rand2009,keeling1999effects,gross2006epidemic,hebert2013pathogen} continue to provide a fruitful framework 
for modelling dynamic or adaptive networks involving epidemics \cite{gross2006epidemic,szabo2014oscillating}, social interactions \cite{demirel2014moment} and ecological 
systems \cite{hebert2013pathogen}. Such models come with the added benefit of some degree of analytical tractability and the means toward 
explicit analytical quantities such as the basic reproduction number and final epidemic size \cite{keeling1999effects}. 
Recently, however there is renewed interest in modelling non-Markovian processes, such as epidemics on 
networks \cite{min2011spreading,cooper2013non,van2013non,jo2014analytically},  
random walks \cite{hoffmann2012generalized} and temporal networks \cite{moinet2014burstiness}. 
This recent burst of research focusing on  non-Markovian dynamics is strongly motivated 
by empirical observations. These show 
that for many real world settings, the Markovian framework is not satisfactory in 
describing temporal statistics, such as time intervals between discrete,
consecutive events. Examples include inter-order and inter-trade durations in financial markets \cite{scalas2006durations}, socio-networks \cite{malmgren2008poissonian}, 
or individual-to-individual contacts being dynamic \cite{moinet2014burstiness}. In the context of epidemiology, 
the period of infectiousness has paramount importance \cite{lloyd2001realistic,distrib}, and researchers departed from the simplifying assumption of exponential distributions by
approximating the empirical distribution of observed infectious periods of various diseases by log-normal and gamma (smallpox \cite{lognormal,gamma}), fixed-length (measles \cite{fixed}) or Weibull distributions (ebola \cite{ebola}). The reliable tools and mathematical machinery of Markovian theory do not translate directly
to modelling and analysis of non-Markovian
systems, and this is the main source of many challenges.

In this letter, we present the first analog 
of pairwise models for non-Markovian epidemics, and show that this is equivalent to 
a set of delay differential equations which (a) show excellent agreement with simulation and (b) allows us to give an implicit 
analytic expression for the final epidemic size, as well as to define 
a new $\mathcal{R}_0$-like quantity which emerges naturally from our calculations.

We consider an undirected and unweighted network with $N$ nodes and an average degree $n$. 
Each node can be susceptible ($S$), infected ($I$) and recovered ($R$). 
For Markovian epidemics, with transmission rate $\tau$ and recovery rate $\gamma$,
the epidemic is well approximated by the pairwise model \cite{keeling1999effects} given below
\begin{eqnarray*}
\dot{[S]}&=&-\tau [SI], \dot{[I]}=\tau [SI]-\gamma [I], \dot{[SS]}=-2\tau [SSI],\\
\dot{[SI]}&=&\tau [SSI]-\tau [ISI]-\tau [SI]-\gamma [SI],
\end{eqnarray*}
where $[X]$, $[XY]$ and $[XYZ]$ are the expected number of nodes in state $X$, links
in state $X-Y$ and triples in state $X-Y-Z$, respectively. Considering the
network at a given time, then counting amounts to $[X]=\sum_{i=1}^{N}X_i$, $[XY]=\sum_{i,j=1}^{N}X_iY_jg_{ij}$ and $[XYZ]=\sum_{i,j,k=1}^{N}X_iY_jZ_kg_{ij}g_{jk}$, where $X, Y, Z \in \{S, I, R\}$, and $G=(g_{ij})_{i,j=1,2,\dots,N}$ is the adjacency matrix of the network such that $g_{ii}=0$, $g_{ij}=g_{ji}$ and $g_{ij}=g_{ji}=1$ if nodes $i$ and $j$ are connected and zero otherwise. Moreover, $X_i$ returns one if node $i$ is in state $X$ and zero otherwise. The dependence on higher order moments can be broken
by using that $[XSY]=\frac{n-1}{n} \frac{[XS] [SY]}{[S]}$ \cite{keeling1999effects}. Applying this leads to the following self-consistent system
\begin{eqnarray}
\label{eq:pairmarkov}
\dot{[S]}&=&-\tau [SI], \dot{[I]}=\tau [SI]-\gamma [I],\nonumber \\
\dot{[SS]}&=&-2\tau \frac{n-1}{n}\frac{[SS][SI]}{[S]},\nonumber \\
\dot{[SI]}&=&\tau \frac{n-1}{n} \frac{[SS][SI]}{[S]}-\tau \frac{n-1}{n} \frac{[SI][SI]}{[S]} \nonumber \\
&-&\tau [SI]-\gamma [SI].
\end{eqnarray}
By applying the closure at the level of pairs, $[XY]=n[X]\frac{[Y]}{N}$, system (\ref{eq:pairmarkov}) reduces to the classic compartmental $SIR$ model,
\begin{equation}
\label{eq:markmeanfield}
\dot{S}=-\tau \frac{n}{N} S I, \dot{I}=\tau \frac{n}{N} S I-\gamma I.
\end{equation}

We wish to apply the previous approach to the case when the recovery time is not exponentially distributed.
First, a fixed infectious period, denoted by $\sigma$, is considered, and the derivation of the pairwise model from first principles is illustrated. 
We show that the non-Markovian dynamics can be described by a delay differential equation with constant delay.
The infection process is assumed to be Markovian, thus the equation for $[S]$ is the same as before, namely $\dot{[S]}(t)=-\tau [SI](t)$.
The number of infected nodes at time $t$ is replenished by $\tau [SI](t)$ and is depleted by $\tau [SI](t-\sigma)$, and this yields
$\dot{[I]}(t)=\tau [SI](t)-\tau [SI](t-\sigma)$. The equation for the number of $S-S$ links is the same because the infection process is Markovian, see (\ref{eq:pairmarkov}).
In a similar manner, the number of $S-I$ links is replenished by $\tau \frac{n-1}{n} \frac{[SS](t)[SI](t)}{[S](t)}$, which is the rate of depletion of $S-S$ links. Furthermore, depletion 
occurs due to the infection within $S-I$ pairs,  $\tau [SI](t)$, and due to the infection of a central $S$ node in an $I-S-I$ triple, $\frac{\tau(n-1)}{n [S](t)} [SI](t)[SI](t)$. On the other 
hand, there are $S-I$ links, which survive the time interval $\sigma$, that will be removed due to the recovery of the $I$ node. However, one needs to account for the removal of $S-I$ links
which were created precisely $\sigma$ times ago. Naively, one would believe that this term is simply proportional to $\tau \frac{n-1}{n}\frac{[SS](t-\sigma)[SI](t-\sigma)}{[S](t-\sigma)}$. 
However, one must account for the fact that in the time interval $(t-\sigma,t)$ an  $S-I$ link could have been destroyed either due to within pair infection or by infection of the $S$ node from outside. Hence, it is obvious that a discount factor needs to be determined to account for this effect.
To calculate this factor, $S-I$ links, that are created at the same time, are considered as a cohort denoted by $x$, and we model infection within and from outside by writing down the following evolution equation,
\begin{align}
x'(t)=-\frac{\tau(n-1)}{n [S](t)} [SI](t)x(t) -\tau x(t), \label{cohort}
\end{align}
where, the first term denotes the `outer' infection of the $S$ node, while the second term stands for `inner' infection of the $S$ node. 
We note that the outside infection is simply proportional to the probability that an $S$ node with an already engaged link has a further susceptible neighbour, $\frac{(n-1)[SI]}{n [S]}$. 
The solution of equation \eqref{cohort} in time interval $[t-\sigma,t]$ is
\begin{align*}
& x(t)=x(t-\sigma) e^{-\int_{t-\sigma}^{t} \left( \frac{\tau(n-1)}{n [S](u)} [SI](u) + \tau\right) du},
\end{align*}
and this provides the depletion or discount rate of $S-I$ links. 
In this case, $x(t-\sigma)=\tau \frac{n-1}{n} \frac{[SS](t-\sigma)[SI](t-\sigma)}{[S](t-\sigma)}$, which is the replenishment of $S-I$ links. Therefore,  summarising all the above, the pairwise DDE for the non-Markovian case is  
%
%
\begin{align}
\label{eq:pairnonmark}
&\dot{[S]}(t)=-\tau [SI](t), \dot{[I]}(t)= \tau [SI](t) - \tau [SI](t-\sigma) \nonumber \\
&\dot{[SS]}(t)=-2\tau \frac{n-1}{n} \frac{[SS](t) [SI](t)}{[S](t)}, \dot{[SI]}(t)= -\tau [SI](t) \nonumber \\
&-\frac{\tau(n-1)}{n [S](t)} [SI](t)[SI](t)+\tau \frac{n-1}{n}\frac{[SS](t)[SI](t)}{[S](t)} \\
&-\tau \frac{n-1}{n}\frac{[SS](t-\sigma)[SI](t-\sigma)}{[S](t-\sigma)} e^{-\int_{t-\sigma}^{t} \left([SI](u)  \frac{\tau(n-1)}{n [S](u)}+\tau \right)du}. \nonumber
\end{align}
This system is now the main subject of our investigation from analytical and numerical point of view.
Similarly to Markovian case, the non-Markovian mean-field model for fixed infectious period is
\begin{eqnarray}
\label{eq:nonmarkmeanfield}
\dot{S}(t)&=&-\tau \frac{n}{N}S(t)I(t),\nonumber \\
\dot{I}(t)&=&\tau \frac{n}{N} S(t)I(t)-\tau \frac{n}{N} S(t-\sigma)I(t-\sigma).
\end{eqnarray}

The most important qualitative results for $SIR$ models are the explicit formula of basic reproduction number and an implicit equation for the final epidemic size. In what follows, we introduce a general concept for the reproduction number associated to pairwise models, and we refer to this as the \textit{pairwise reproduction number}. Using this concept, the final size relations for the above mean-field, classic pairwise and DDE-based pairwise models are derived. Reproduction numbers play a crucial role in mathematical epidemiology, so we begin by investigating these. The basic reproduction number $\mathcal{R}_{0}$ denotes the expected number of secondary infections caused by a `typical' infected individual during its infectious period when placed in a fully susceptible population, which is a definition understood at the level of nodes (individuals). On the other hand, the pairwise model is written at the level of links and describes the dynamics of susceptible ($S-S$) and infected ($S-I$) links. This fact gives us an opportunity to define a new type of reproduction numbers, which we call \textit{pairwise reproduction number} and denote it by $\mathcal{R}^{p}_{0}$. More precisely, we distinguish the following two useful quantities: (a) the \textit{basic} reproduction number is the expected lifetime of an $I$ \textbf{node} multiplied by the number of newly infected \textbf{nodes} per unit time, and (b) the \textit{pairwise} reproduction number is the expected lifetime of an $S-I$ \textbf{link} multiplied by the number of newly generated $S-I$ \textbf{links} per unit time.

An infected node is removed due to its recovery, thus in general the expected lifetime is the expected value of a random variable $X$ corresponding to the distribution of the length of infectious periods. In contrast, an $S-I$ link can be removed due to the recovery of the $I$ node but also due to the infection of the $S$ node. Therefore, the expected lifetime of the $S-I$ link is the expected value of the minimum of two random variables. If we assume that the process of infection along such a link has density function $f_i$ with survival function $\xi_i$, and the process of recovery has density function $f_r$ with survival function $\xi_r$, then, denoting by $Z$ the random variable defined by the lifetime of an $S-I$ link, we have
\begin{equation}
	\label{eq:lifetime}
	\mathbb{E}(Z)=\int_{-\infty}^{\infty} x \left(f_i(x) \xi_r(x)+f_r(x) \xi_i(x)\right) dx.
\end{equation}

From the assumption that the infection time along $S-I$ links is exponentially distributed, the number of newly infected nodes per unit time is $\frac{n}{N}\tau [S]_0$ in the mean-field model, and the expected number of newly infected links is $\tau \frac{n-1}{n}\frac{[SS]_0}{[S]_0}=\tau \frac{n-1}{N}{[S]_0}$ in the pairwise model, where we used the approximation $[SS]_0=\frac{n}{N}[S]^2_{0}$. 

We illustrate how to use the formula \eqref{eq:lifetime} in the case of fixed length infectious period ($\sigma$). In this case, the survival function is
\[
\xi_r(t)= \begin{cases}
1 & \textrm{if $0\leq t<\sigma$,} \\
0 & \textrm{if $t\geq\sigma$}, \\	
\end{cases}
\]
and the density function $f_r(t)$ is the Dirac-delta $\delta(t-\sigma)$. Using fundamental properties of the delta function, we have 
\begin{eqnarray*}
	\mathbb{E}(Z)&=&\int_{-\infty}^{\infty} x f_i(x) \xi_r(x) dx + \int_{-\infty}^{\infty} x f_r(x) \xi_i(x) dx \nonumber\\
	&=&\left(-\sigma e^{-\tau \sigma}+\frac{1-e^{-\tau \sigma}}{\tau}\right)+\sigma e^{-\tau \sigma},
\end{eqnarray*} 
and multiplying this result by the number of newly generated $S-I$ links, the formula in Table \ref{R0_table} for $\mathcal{R}_0^p$ follows.
More importantly, it is noteworthy to highlight the general result that $\mathbb{E}(Z)$, upon using that the infection process is Markovian, reduces to 
evaluating the Laplace transform of the density of the recovery time. This provides a very general result, which in many cases leads to an explicit analytical result for $\mathcal{R}^p_0$, see Table \ref{R0_table}.
%
     \begin{table}[Htbp]
     \begin{ruledtabular}
     \begin{tabular}{c c c }
		  & $\mathcal{R}_0$ & $\mathcal{R}^p_0$ \\
		Markovian & $\frac{n}{N}\frac{\tau}{\gamma}S_{0} $ & $\frac{n-1}{N}\frac{\tau}{\tau+\gamma}[S]_{0} $ \\
		Fixed & $\frac{n}{N}\tau \sigma S_{0} $ & $\frac{n-1}{N}(1-e^{-\tau \sigma})[S]_{0} $ \\
		General &  $\frac{n}{N}\tau \mathbb{E}(X) S_{0}$ & $\frac{n-1}{N}\left(1-\mathcal L[f_r](\tau)\right)[S]_0$ \\ 
     \end{tabular}
     \end{ruledtabular}
		\caption{Basic and pairwise reproduction numbers for different recovery distributions. $\mathcal L[f_r](\tau)$ denotes the Laplace transform of $f_r$, the density of the recovery process, at $\tau$.}
\label{R0_table}
     \end{table}

For the standard Markovian mean-field model, the process of calculating the final epidemic size is well-known.
From Eq. (\ref{eq:markmeanfield}), we evaluate $d I/d S$ and integrate it to obtain
\begin{equation*}
\ln\left(\frac{S_{\infty}}{S_0}\right)=\frac{\tau}{\gamma}\frac{n}{N}S_0 \left(\frac{S_{\infty}}{S_0}-1\right).
\end{equation*}
Using that $\mathcal{R}_0=\frac{\tau}{\gamma}\frac{n}{N}S_0$, we have
\begin{equation}
\label{eq:standardfs}
\ln\left(\frac{S_{\infty}}{S_0}\right)=\mathcal{R}_0 \left(\frac{S_{\infty}}{S_0}-1\right).
\end{equation}
The final epidemic size (i.e. the total number of infections) can be easily computed by using $R_{\infty}=N-S_{\infty}$. In the non-Markovian case, the calculations (which are included in the supplemental material) are rather different and the resulting final size relation is 
\vspace*{-0.6ex}
\begin{equation}
\label{eq:meannonmarkov}
\ln\left(\frac{S_{\infty}}{S_0}\right)=\tau \frac{n}{N} \sigma S_0 \left(\frac{S_{\infty}}{S_0}-1\right).
\end{equation}
 As in this case $\mathcal{R}_0=\tau\frac{n}{N} \sigma S_0$, the final size relation (\ref{eq:meannonmarkov}) shows the `standard' form of (\ref{eq:standardfs}).
\label{pairwise}
The dynamical systems  \eqref{eq:pairmarkov} and \eqref{eq:pairnonmark} can be manipulated conveniently to derive an analytic relation between the final epidemic size and the basic reproduction number.  This is known for the Markovian case but it is a new result for the non-Markovian one.
While the full derivation for the non-Markovian case is given in the supplemental material, the main steps of the calculations are: (a) find an invariant to reduce the dimensionality of the system, (b) integrate the equation for $[SI](t)$, (c) integrate the equation for $[S](t)$ on $[0,\infty)$ and (d) employ algebraic manipulations to obtain the final size relation. Following this procedure yields
\begin{eqnarray}
\label{eq:pairnonmarkov}
\frac{s_\infty^{\frac{1}{n}}-1}{\frac{1}{n-1}}=\frac{n-1}{N}\left(1-e^{-\tau \sigma}\right)[S]_0\left(s_\infty^{\frac{n-1}{n}}-1\right),
\end{eqnarray}
where $s_{\infty}=\frac{[S]_\infty}{[S]_0}$ and the attack rate is simply $1-s_\infty$. Using the same technique for the Markovian case leads to
\begin{eqnarray}
\label{eq:markovpair}
\frac{s_\infty^{\frac{1}{n}}-1}{\frac{1}{n-1}}=\frac{n-1}{N}\frac{\tau}{\tau+\gamma}[S]_0
\left(s_\infty^{\frac{n-1}{n}}-1\right).
\end{eqnarray}
Upon inspecting the two relations above, the following important observations can be made. 
First, the implicit relation between final size and $\mathcal{R}^p_0$ is conserved between the Markovian and non-Markovian model.
Moreover, upon using the values of $\mathcal{R}^p_0$ as given in Table \ref{R0_table}, equations \eqref{eq:pairnonmarkov} and \eqref{eq:markovpair} can be cast in the following general form
\begin{eqnarray}
\label{eq:standardpfs}
\frac{s_\infty^{\frac{1}{n}}-1}{\frac{1}{n-1}}=\mathcal{R}^p_0\left(s_\infty^{\frac{n-1}{n}}-1\right).
\end{eqnarray}
In fact we conjecture that this relation will hold true for pairwise models with different infectivity period profiles.
The second observation is that taking the limit of $n \rightarrow \infty$ in \eqref{eq:standardpfs} gives rise to 
\begin{equation}
\label{eq:sinffs}
\ln (s_\infty) = \mathcal{R}^p_0 (s_\infty -1),
\end{equation}
which is equivalent to the `standard' form of (\ref{eq:standardfs})
%
%

\begin{figure*}
	\centering
	\begin{minipage}[b]{.45\textwidth}
		\includegraphics[width=\linewidth]{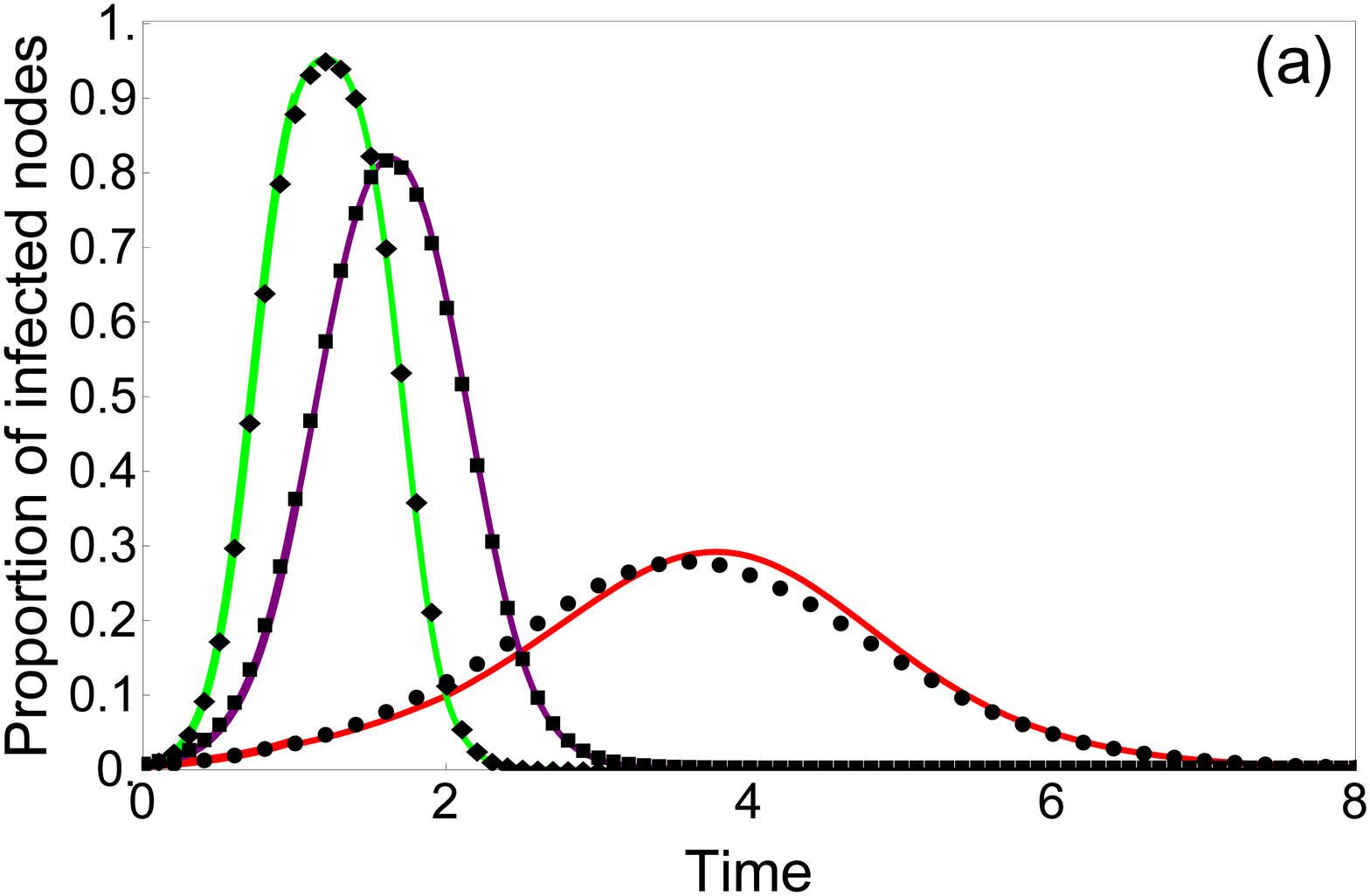}
		\includegraphics[width=\linewidth]{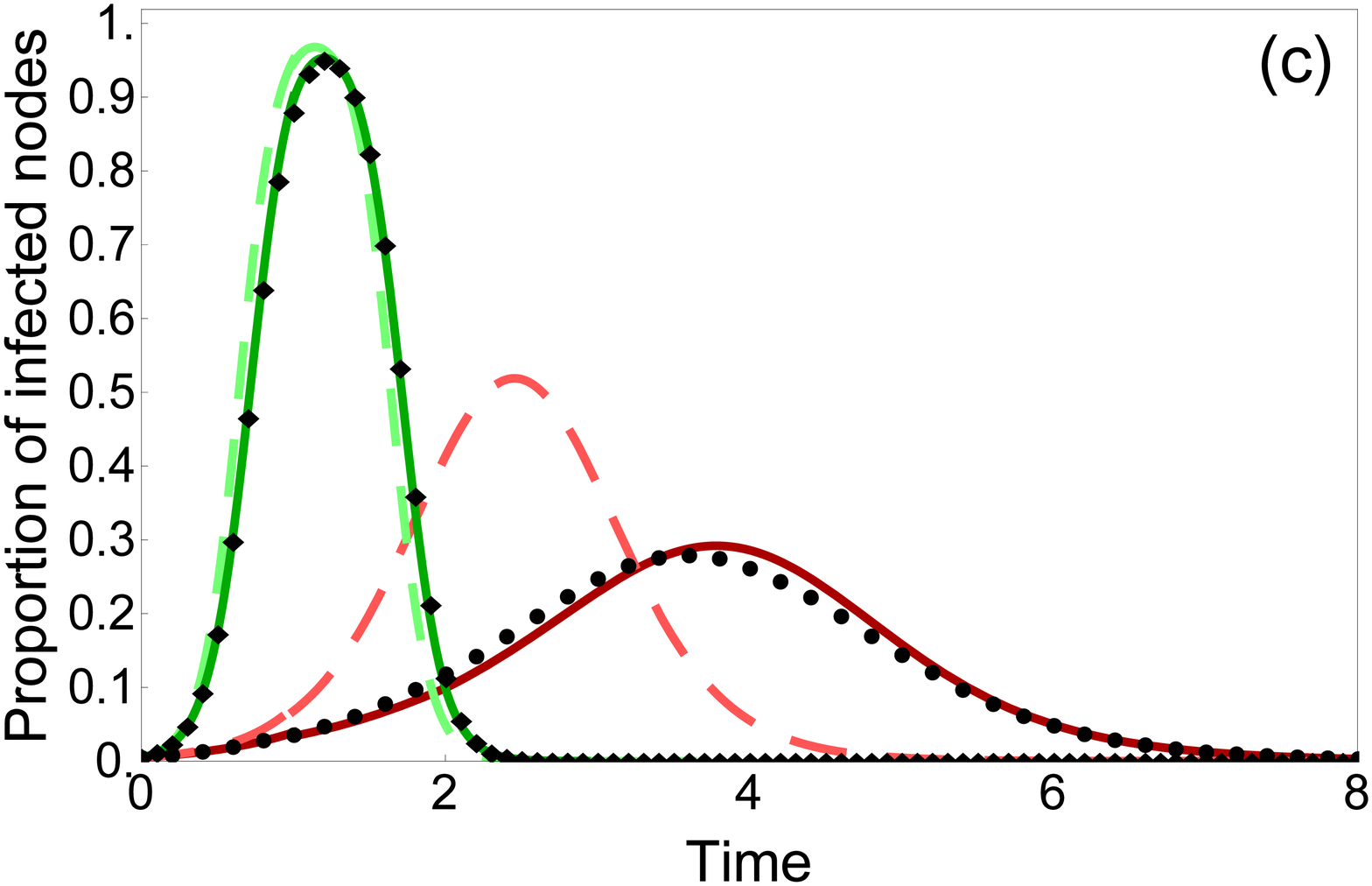}		
	\end{minipage}
	\begin{minipage}[b]{.45\textwidth}
		\includegraphics[width=\linewidth]{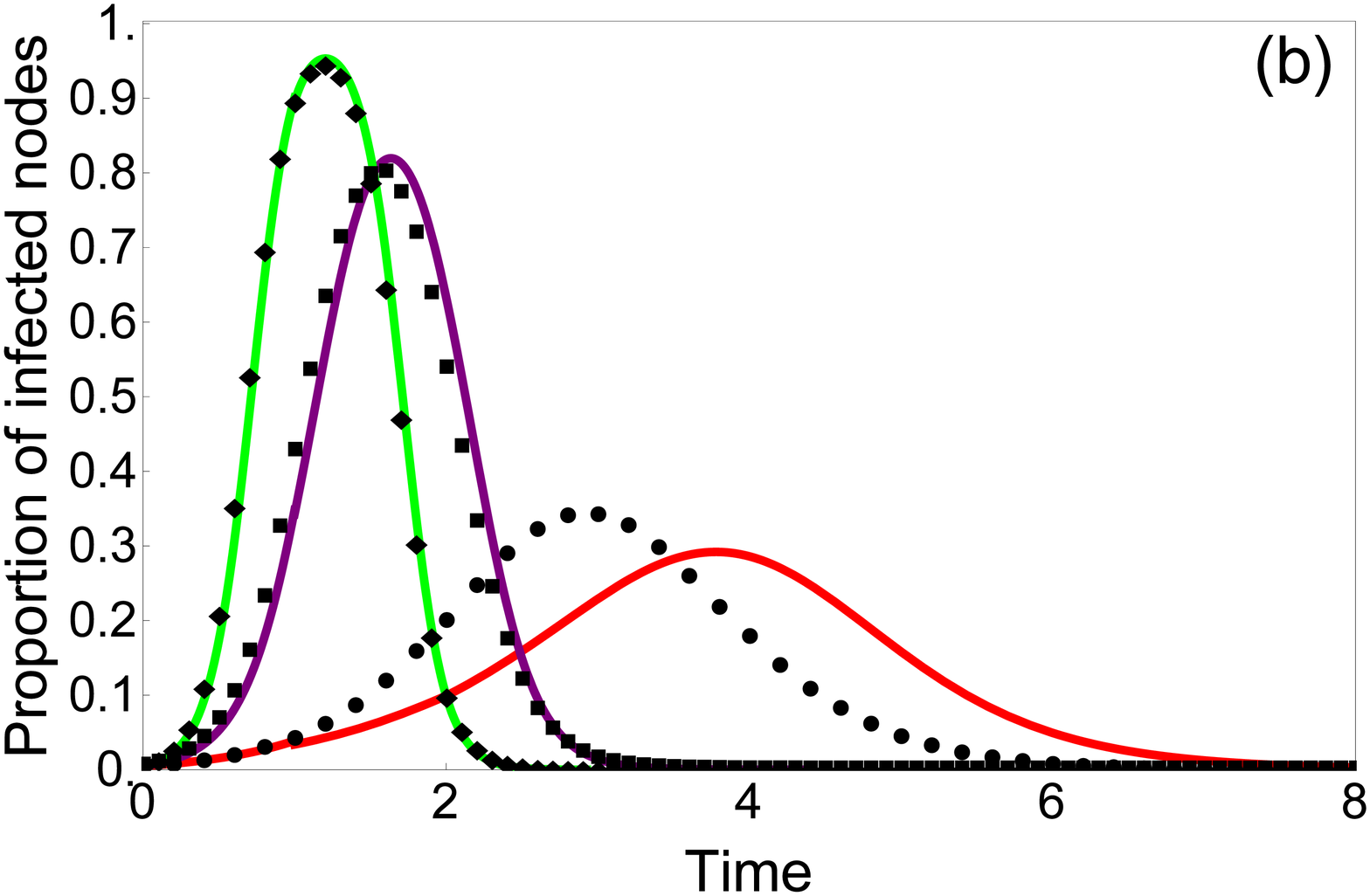}		
		\topinset{\includegraphics[height=2.8cm]{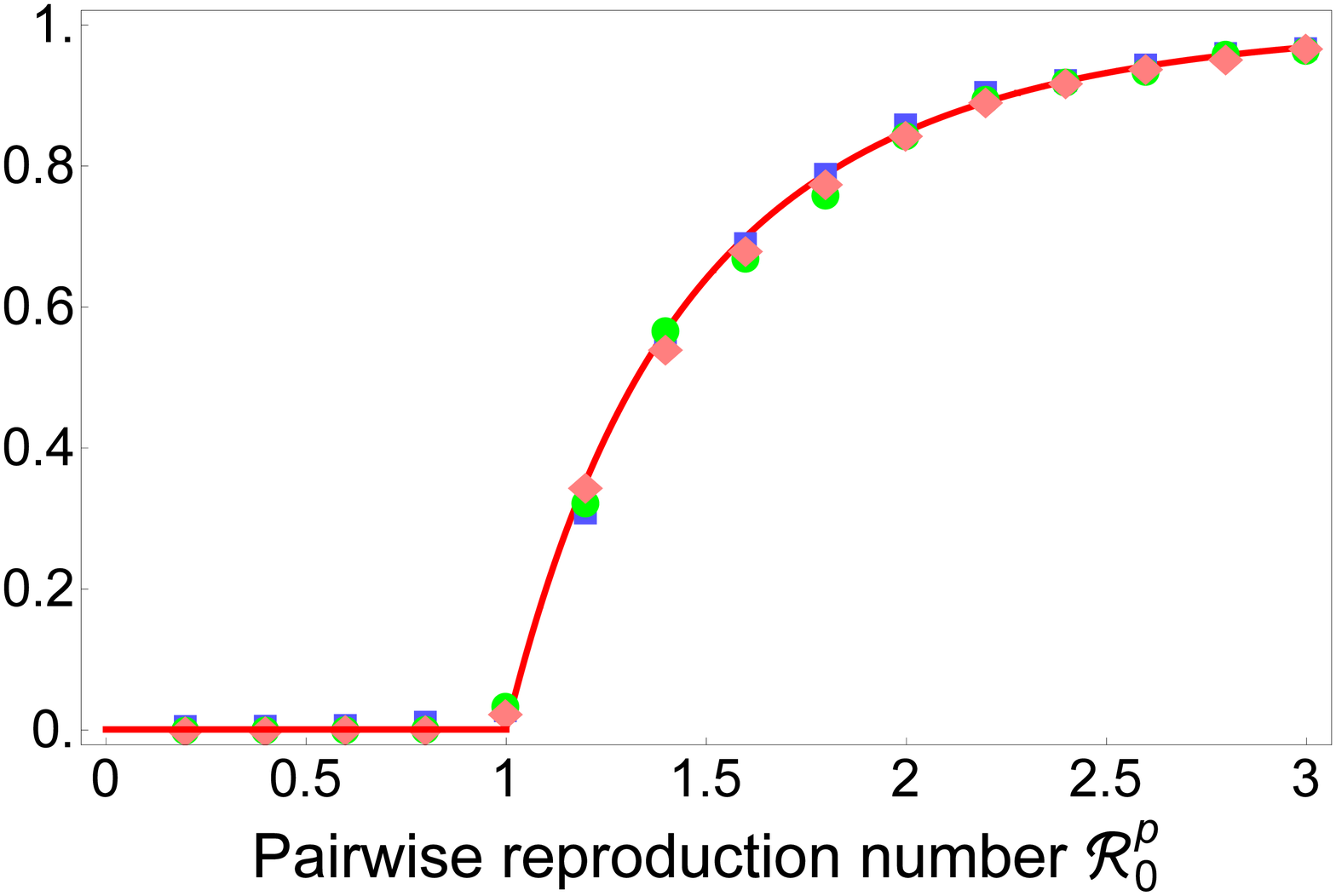}}{\includegraphics[height=5cm]{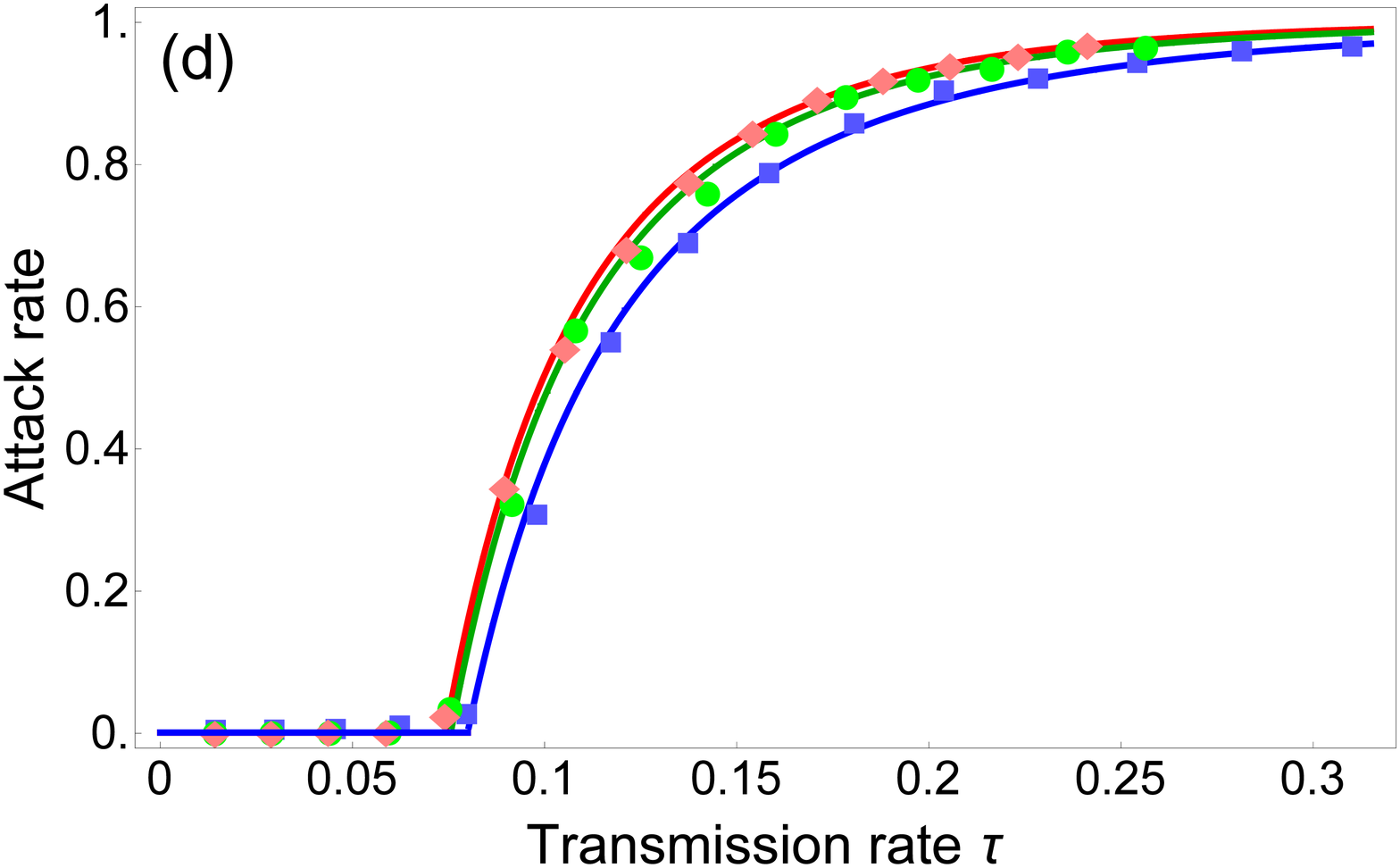}}{45pt}{51pt}
	\end{minipage}
	\caption{Simulations of non-Markovian epidemics on networks with $N=1000$ nodes:  (a) solid lines show the numerical solution of (\ref{eq:pairnonmark}) and the circles/squares/diamonds correspond to simulations for homogeneous networks with $\langle k\rangle=5/10/15$, respectively;  (b) the same as before but for Erd\H os-R\'enyi random networks with $\langle k\rangle=5/10/15$;  (c) the solid and dashed lines show the numerical solution of pairwise (\ref{eq:pairnonmark}) and mean-field (\ref{eq:nonmarkmeanfield}) models, respectively and, for homogeneous networks with $\langle k\rangle=5$ and $\langle k\rangle=15$. For (a), (b) and (c) the transmission rate is $\tau=0.55$ and the infectious period is fixed, $\sigma=1$. Finally, (d) the diamonds/circles/squares correspond to numerical simulations using homogeneous network with $\langle k\rangle=15$ and using fixed and two different but gamma distributed infectious periods ($\circ$ - shape $\alpha=2$, scale $\beta=\frac{1}{2}$, $\square$ - shape $\alpha=\frac{1}{2}$, scale $\beta=2$), respectively. The solid lines correspond to the analytical final epidemic size for fixed (\ref{eq:pairnonmarkov}) and general (\ref{eq:geninffinalsize}) infectious periods. 
	The inset shows the analytical and the simulated final epidemic sizes plotted against the pairwise reproduction number.}
\label{fig:1}
\end{figure*}

To test the validity of our model we implemented an algorithm to simulate the non-Markovian SIR process with arbitrary recovery times, and considered random networks with $N=1000$ nodes. In Fig.~\ref{fig:1}(a,b) homogenous and Erd\H{o}s-R\'enyi random networks are considered, respectively. Here, the mean of 100 simulations is compared to the solution of system \eqref{eq:pairnonmark}. The agreement is excellent for homogenous networks even for low degrees. Despite the pairwise model not accounting explicitly for the network's degree distribution, the agreement is surprisingly good for relatively dense Erd\H{o}s-R\'enyi networks.


In Fig.~\ref{fig:1}(c) we compare and contrast the differences between simulations, mean-field and pairwise models for the non-Markovian case. 
For denser networks ($\langle k\rangle=15$), both models perform well with the pairwise yielding a better agreement with output from simulation.
However, the difference is striking for sparser networks ($\langle k\rangle=5$), where the mean-field approximation performs very poorly, while the pairwise DDE model leads to good agreement even in this case.
%

In Fig.~\ref{fig:1} (d), analytic final size relations are tested against simulation results for a range of different infectious period distributions, all sharing the same mean. The horizontal axis corresponds to the transmission rate $\tau$, and the plot highlights the threshold dynamics, as well as the necessity to correctly model the recovery time distribution in order to avoid under or over estimation of the final epidemic size. Based on Table \ref{R0_table}, the analytical expressions for $\mathcal{R}^p_0$ can be computed for the Markovian (or exponential), fixed  and gamma-distributed recovery times. These values are
\begin{eqnarray*}
&\mathcal{R}&^p_{0,\mathrm{\Gamma}(\frac{1}{2},2)}=c\left(1-\frac{1}{\sqrt{1+2 \tau}}\right), \mathcal{R}^p_{0,\mathrm{Exp}(1)}=c\left(\frac{\tau}{\tau+1}\right),\\
&\mathcal{R}&^p_{0,\mathrm{\Gamma}(2,\frac{1}{2})}=c\left(1-\frac{4}{(2+\tau)^2}\right), \mathcal{R}^p_{0,\mathrm{Fixed}(1)}=c \left(1-e^{-\tau}\right),
\end{eqnarray*}
where $c=\frac{(n-1)[S]_0}{N}$, and satisfy the following inequality
\begin{equation}
\mathcal{R}^p_{0,\mathrm{\Gamma}(\frac{1}{2},2)}\leq \mathcal{R}^p_{0,\mathrm{Exp}(1)}\leq\mathcal{R}^p_{0,\mathrm{\Gamma}(2,\frac{1}{2})}\leq \mathcal{R}^p_{0,\mathrm{Fixed}(1)}.
\end{equation}
We note that (a) all recovery time distributions have the same mean $1$ and (b) the variances satisfy the converse inequality, with higher variance in recovery time (i.e. 2, 1, 1/2 and 0) giving a smaller $\mathcal{R}^p_0$ value, despite $\tau$ being fixed. The overall agreement between the analytic results of the pairwise model and the stochastic simulations is excellent and confirms the validity of our 
final size relations.  The inset in Fig.~1 (d) illustrates how the final epidemic size depends on the pairwise reproduction number, and shows that the same value of $\mathcal{R}^p_0$ produces the same attack rate, regardless of the distribution from where it is originated from, in accordance with our formula \eqref{eq:standardpfs}.


We have introduced a generalization of pairwise models to non-Markovian epidemics with fixed infectious period. The resulting model is a system of delay differential equations with constant delay and we have provided as a full as possible analytical and numerical analysis of this model and benchmarked its performance against explicit stochastic network simulations. We have presented a new concept of reproduction numbers introducing the \textit{pairwise} reproduction number $\mathcal{R}^p_0$ and have derived the final epidemic size relation for non-Markovian mean-field and pairwise DDE models. 

The numerical solution of the non-Markovian pairwise DDE shows excellent agreement with results based on explicit stochastic network simulations and sheds some light on the impact of non-Markovianity. More importantly, via the analytic results we can gain insight how and where non-Markovianity enters and impacts upon important epidemic descriptors.

The model and results in this paper should provide a framework for deeper and more comprehensive analysis of non-Markovian processes on networks and these should not be restricted to epidemics with fixed delays. Preliminary investigations indicate that our model can be extended to consider arbitrary recovery time distributions. In this case, the resulting model is a more complex integro-differential equation requiring a more challenging and elaborate analysis. Nevertheless, it turns out that the final epidemic size relation, upon assuming a general probability distribution with density function ($f_r$), yields 
\begin{equation}
\label{eq:geninffinalsize}
\frac{s_\infty^{\frac{1}{n}}-1}{\frac{1}{n-1}}=\frac{n-1}{N}\left(1-\mathcal L[f_r](\tau)\right)[S]_0
\left(s_\infty^{\frac{n-1}{n}}-1\right),
\end{equation}
which agrees with the general equation suggested in \eqref{eq:standardpfs}. For recovery of fixed length, relation \eqref{eq:geninffinalsize} reduces to \eqref{eq:pairnonmarkov}. The validity of our general final size relation \eqref{eq:geninffinalsize} was tested also for different gamma distributions, see Fig.~1(d), showing a strong predictive power for general non-Markovian epidemics on networks. The difficulty of modelling non-Markovian processes is well known, but our current framework can pave the way for identifying fruitful links between different areas of delay differential equations, stochastic processes, dynamical systems and epidemiological models. 
\section{References}
\bibliography{Zs_G_I}

\begin{thebibliography}{25}%
\makeatletter
\providecommand \@ifxundefined [1]{%
 \@ifx{#1\undefined}
}%
\providecommand \@ifnum [1]{%
 \ifnum #1\expandafter \@firstoftwo
 \else \expandafter \@secondoftwo
 \fi
}%
\providecommand \@ifx [1]{%
 \ifx #1\expandafter \@firstoftwo
 \else \expandafter \@secondoftwo
 \fi
}%
\providecommand \natexlab [1]{#1}%
\providecommand \enquote  [1]{``#1''}%
\providecommand \bibnamefont  [1]{#1}%
\providecommand \bibfnamefont [1]{#1}%
\providecommand \citenamefont [1]{#1}%
\providecommand \href@noop [0]{\@secondoftwo}%
\providecommand \href [0]{\begingroup \@sanitize@url \@href}%
\providecommand \@href[1]{\@@startlink{#1}\@@href}%
\providecommand \@@href[1]{\endgroup#1\@@endlink}%
\providecommand \@sanitize@url [0]{\catcode `\\12\catcode `\$12\catcode
  `\&12\catcode `\#12\catcode `\^12\catcode `\_12\catcode `\%12\relax}%
\providecommand \@@startlink[1]{}%
\providecommand \@@endlink[0]{}%
\providecommand \url  [0]{\begingroup\@sanitize@url \@url }%
\providecommand \@url [1]{\endgroup\@href {#1}{\urlprefix }}%
\providecommand \urlprefix  [0]{URL }%
\providecommand \Eprint [0]{\href }%
\providecommand \doibase [0]{http://dx.doi.org/}%
\providecommand \selectlanguage [0]{\@gobble}%
\providecommand \bibinfo  [0]{\@secondoftwo}%
\providecommand \bibfield  [0]{\@secondoftwo}%
\providecommand \translation [1]{[#1]}%
\providecommand \BibitemOpen [0]{}%
\providecommand \bibitemStop [0]{}%
\providecommand \bibitemNoStop [0]{.\EOS\space}%
\providecommand \EOS [0]{\spacefactor3000\relax}%
\providecommand \BibitemShut  [1]{\csname bibitem#1\endcsname}%
\let\auto@bib@innerbib\@empty
\bibitem [{\citenamefont {Newman}(2003)}]{newman2003structure}%
  \BibitemOpen
  \bibfield  {author} {\bibinfo {author} {\bibfnamefont {M.~E.}\ \bibnamefont
  {Newman}},\ }\href@noop {} {\bibfield  {journal} {\bibinfo  {journal} {SIAM
  review}\ }\textbf {\bibinfo {volume} {45}},\ \bibinfo {pages} {167} (\bibinfo
  {year} {2003})}\BibitemShut {NoStop}%
\bibitem [{\citenamefont {Boccaletti}\ \emph {et~al.}(2006)\citenamefont
  {Boccaletti}, \citenamefont {Latora}, \citenamefont {Moreno}, \citenamefont
  {Chavez},\ and\ \citenamefont {Hwang}}]{boccaletti2006complex}%
  \BibitemOpen
  \bibfield  {author} {\bibinfo {author} {\bibfnamefont {S.}~\bibnamefont
  {Boccaletti}}, \bibinfo {author} {\bibfnamefont {V.}~\bibnamefont {Latora}},
  \bibinfo {author} {\bibfnamefont {Y.}~\bibnamefont {Moreno}}, \bibinfo
  {author} {\bibfnamefont {M.}~\bibnamefont {Chavez}}, \ and\ \bibinfo {author}
  {\bibfnamefont {D.-U.}\ \bibnamefont {Hwang}},\ }\href@noop {} {\bibfield
  {journal} {\bibinfo  {journal} {Physics reports}\ }\textbf {\bibinfo {volume}
  {424}},\ \bibinfo {pages} {175} (\bibinfo {year} {2006})}\BibitemShut
  {NoStop}%
\bibitem [{\citenamefont {Pastor-Satorras}\ \emph {et~al.}(2014)\citenamefont
  {Pastor-Satorras}, \citenamefont {Castellano}, \citenamefont {Van~Mieghem},\
  and\ \citenamefont {Vespignani}}]{pastor2014epidemic}%
  \BibitemOpen
  \bibfield  {author} {\bibinfo {author} {\bibfnamefont {R.}~\bibnamefont
  {Pastor-Satorras}}, \bibinfo {author} {\bibfnamefont {C.}~\bibnamefont
  {Castellano}}, \bibinfo {author} {\bibfnamefont {P.}~\bibnamefont
  {Van~Mieghem}}, \ and\ \bibinfo {author} {\bibfnamefont {A.}~\bibnamefont
  {Vespignani}},\ }\href@noop {} {\bibfield  {journal} {\bibinfo  {journal}
  {arXiv preprint arXiv:1408.2701}\ } (\bibinfo {year} {2014})}\BibitemShut
  {NoStop}%
\bibitem [{\citenamefont {Pastor-Satorras}\ and\ \citenamefont
  {Vespignani}(2001)}]{SatorrasVespignani}%
  \BibitemOpen
  \bibfield  {author} {\bibinfo {author} {\bibfnamefont {R.}~\bibnamefont
  {Pastor-Satorras}}\ and\ \bibinfo {author} {\bibfnamefont {A.}~\bibnamefont
  {Vespignani}},\ }\href@noop {} {\bibfield  {journal} {\bibinfo  {journal}
  {Phys. Rev. Lett.}\ }\textbf {\bibinfo {volume} {86}},\ \bibinfo {pages}
  {3200} (\bibinfo {year} {2001})}\BibitemShut {NoStop}%
\bibitem [{\citenamefont {Liu}\ \emph {et~al.}(2013)\citenamefont {Liu},
  \citenamefont {R\"{o}st},\ and\ \citenamefont {Vas}}]{liurostvas}%
  \BibitemOpen
  \bibfield  {author} {\bibinfo {author} {\bibfnamefont {M.}~\bibnamefont
  {Liu}}, \bibinfo {author} {\bibfnamefont {G.}~\bibnamefont {R\"{o}st}}, \
  and\ \bibinfo {author} {\bibfnamefont {G.}~\bibnamefont {Vas}},\ }\href@noop
  {} {\bibfield  {journal} {\bibinfo  {journal} {Comput. Math. Appl.}\ }\textbf
  {\bibinfo {volume} {66}},\ \bibinfo {pages} {1534} (\bibinfo {year}
  {2013})}\BibitemShut {NoStop}%
\bibitem [{\citenamefont {Keeling}(1999)}]{keeling1999effects}%
  \BibitemOpen
  \bibfield  {author} {\bibinfo {author} {\bibfnamefont {M.~J.}\ \bibnamefont
  {Keeling}},\ }\href@noop {} {\bibfield  {journal} {\bibinfo  {journal} {Proc.
  R. Soc. B}\ }\textbf {\bibinfo {volume} {266}},\ \bibinfo {pages} {859}
  (\bibinfo {year} {1999})}\BibitemShut {NoStop}%
\bibitem [{\citenamefont {Rand}(2009)}]{rand2009}%
  \BibitemOpen
  \bibfield  {author} {\bibinfo {author} {\bibfnamefont {D.}~\bibnamefont
  {Rand}},\ }\enquote {\bibinfo {title} {Advanced ecological theory: principles
  and applications},}\ \ (\bibinfo  {publisher} {John Wiley \& Sons},\ \bibinfo
  {year} {2009})\ Chap.\ \bibinfo {chapter} {Correlation equations and pair
  approximations for spatial ecologies}, pp.\ \bibinfo {pages}
  {100--142}\BibitemShut {NoStop}%
\bibitem [{\citenamefont {Gross}\ \emph {et~al.}(2006)\citenamefont {Gross},
  \citenamefont {DÕLima},\ and\ \citenamefont {Blasius}}]{gross2006epidemic}%
  \BibitemOpen
  \bibfield  {author} {\bibinfo {author} {\bibfnamefont {T.}~\bibnamefont
  {Gross}}, \bibinfo {author} {\bibfnamefont {C.~J.~D.}\ \bibnamefont
  {DÕLima}}, \ and\ \bibinfo {author} {\bibfnamefont {B.}~\bibnamefont
  {Blasius}},\ }\href@noop {} {\bibfield  {journal} {\bibinfo  {journal} {Phys.
  Rev. Lett.}\ }\textbf {\bibinfo {volume} {96}},\ \bibinfo {pages} {208701}
  (\bibinfo {year} {2006})}\BibitemShut {NoStop}%
\bibitem [{\citenamefont {H{\'e}bert-Dufresne}\ \emph
  {et~al.}(2013)\citenamefont {H{\'e}bert-Dufresne}, \citenamefont
  {Patterson-Lomba}, \citenamefont {Goerg},\ and\ \citenamefont
  {Althouse}}]{hebert2013pathogen}%
  \BibitemOpen
  \bibfield  {author} {\bibinfo {author} {\bibfnamefont {L.}~\bibnamefont
  {H{\'e}bert-Dufresne}}, \bibinfo {author} {\bibfnamefont {O.}~\bibnamefont
  {Patterson-Lomba}}, \bibinfo {author} {\bibfnamefont {G.~M.}\ \bibnamefont
  {Goerg}}, \ and\ \bibinfo {author} {\bibfnamefont {B.~M.}\ \bibnamefont
  {Althouse}},\ }\href@noop {} {\bibfield  {journal} {\bibinfo  {journal}
  {Phys. Rev. Lett.}\ }\textbf {\bibinfo {volume} {110}},\ \bibinfo {pages}
  {108103} (\bibinfo {year} {2013})}\BibitemShut {NoStop}%
\bibitem [{\citenamefont {Szab{\'o}-Solticzky}\ \emph
  {et~al.}(2014)\citenamefont {Szab{\'o}-Solticzky}, \citenamefont {Berthouze},
  \citenamefont {Kiss},\ and\ \citenamefont {Simon}}]{szabo2014oscillating}%
  \BibitemOpen
  \bibfield  {author} {\bibinfo {author} {\bibfnamefont {A.}~\bibnamefont
  {Szab{\'o}-Solticzky}}, \bibinfo {author} {\bibfnamefont {L.}~\bibnamefont
  {Berthouze}}, \bibinfo {author} {\bibfnamefont {I.~Z.}\ \bibnamefont {Kiss}},
  \ and\ \bibinfo {author} {\bibfnamefont {P.~L.}\ \bibnamefont {Simon}},\
  }\href@noop {} {\bibfield  {journal} {\bibinfo  {journal} {arXiv preprint
  arXiv:1410.4953}\ } (\bibinfo {year} {2014})}\BibitemShut {NoStop}%
\bibitem [{\citenamefont {Demirel}\ \emph {et~al.}(2014)\citenamefont
  {Demirel}, \citenamefont {Vazquez}, \citenamefont {B{\"o}hme},\ and\
  \citenamefont {Gross}}]{demirel2014moment}%
  \BibitemOpen
  \bibfield  {author} {\bibinfo {author} {\bibfnamefont {G.}~\bibnamefont
  {Demirel}}, \bibinfo {author} {\bibfnamefont {F.}~\bibnamefont {Vazquez}},
  \bibinfo {author} {\bibfnamefont {G.}~\bibnamefont {B{\"o}hme}}, \ and\
  \bibinfo {author} {\bibfnamefont {T.}~\bibnamefont {Gross}},\ }\href@noop {}
  {\bibfield  {journal} {\bibinfo  {journal} {Physica D}\ }\textbf {\bibinfo
  {volume} {267}},\ \bibinfo {pages} {68} (\bibinfo {year} {2014})}\BibitemShut
  {NoStop}%
\bibitem [{\citenamefont {Min}\ \emph {et~al.}(2011)\citenamefont {Min},
  \citenamefont {Goh},\ and\ \citenamefont {Vazquez}}]{min2011spreading}%
  \BibitemOpen
  \bibfield  {author} {\bibinfo {author} {\bibfnamefont {B.}~\bibnamefont
  {Min}}, \bibinfo {author} {\bibfnamefont {K.-I.}\ \bibnamefont {Goh}}, \ and\
  \bibinfo {author} {\bibfnamefont {A.}~\bibnamefont {Vazquez}},\ }\href@noop
  {} {\bibfield  {journal} {\bibinfo  {journal} {Phys. Rev. E}\ }\textbf
  {\bibinfo {volume} {83}},\ \bibinfo {pages} {036102} (\bibinfo {year}
  {2011})}\BibitemShut {NoStop}%
\bibitem [{\citenamefont {Cooper}(2013)}]{cooper2013non}%
  \BibitemOpen
  \bibfield  {author} {\bibinfo {author} {\bibfnamefont {F.}~\bibnamefont
  {Cooper}},\ }\href@noop {} {\bibfield  {journal} {\bibinfo  {journal}
  {http://www.dtc.ox.ac.uk/people/13/cooperf
  /files/MA469ThesisFergusCooper.pdf}\ } (\bibinfo {year} {2013})}\BibitemShut
  {NoStop}%
\bibitem [{\citenamefont {Van~Mieghem}\ and\ \citenamefont {Van~de
  Bovenkamp}(2013)}]{van2013non}%
  \BibitemOpen
  \bibfield  {author} {\bibinfo {author} {\bibfnamefont {P.}~\bibnamefont
  {Van~Mieghem}}\ and\ \bibinfo {author} {\bibfnamefont {R.}~\bibnamefont
  {Van~de Bovenkamp}},\ }\href@noop {} {\bibfield  {journal} {\bibinfo
  {journal} {Phys. Rev. Lett.}\ }\textbf {\bibinfo {volume} {110}},\ \bibinfo
  {pages} {108701} (\bibinfo {year} {2013})}\BibitemShut {NoStop}%
\bibitem [{\citenamefont {Jo}\ \emph {et~al.}(2014)\citenamefont {Jo},
  \citenamefont {Perotti}, \citenamefont {Kaski},\ and\ \citenamefont
  {Kert{\'e}sz}}]{jo2014analytically}%
  \BibitemOpen
  \bibfield  {author} {\bibinfo {author} {\bibfnamefont {H.-H.}\ \bibnamefont
  {Jo}}, \bibinfo {author} {\bibfnamefont {J.~I.}\ \bibnamefont {Perotti}},
  \bibinfo {author} {\bibfnamefont {K.}~\bibnamefont {Kaski}}, \ and\ \bibinfo
  {author} {\bibfnamefont {J.}~\bibnamefont {Kert{\'e}sz}},\ }\href@noop {}
  {\bibfield  {journal} {\bibinfo  {journal} {Phys. Rev. X}\ }\textbf {\bibinfo
  {volume} {4}},\ \bibinfo {pages} {011041} (\bibinfo {year}
  {2014})}\BibitemShut {NoStop}%
\bibitem [{\citenamefont {Hoffmann}\ \emph {et~al.}(2012)\citenamefont
  {Hoffmann}, \citenamefont {Porter},\ and\ \citenamefont
  {Lambiotte}}]{hoffmann2012generalized}%
  \BibitemOpen
  \bibfield  {author} {\bibinfo {author} {\bibfnamefont {T.}~\bibnamefont
  {Hoffmann}}, \bibinfo {author} {\bibfnamefont {M.~A.}\ \bibnamefont
  {Porter}}, \ and\ \bibinfo {author} {\bibfnamefont {R.}~\bibnamefont
  {Lambiotte}},\ }\href@noop {} {\bibfield  {journal} {\bibinfo  {journal}
  {Phys. Rev. E}\ }\textbf {\bibinfo {volume} {86}},\ \bibinfo {pages} {046102}
  (\bibinfo {year} {2012})}\BibitemShut {NoStop}%
\bibitem [{\citenamefont {Moinet}\ \emph {et~al.}(2014)\citenamefont {Moinet},
  \citenamefont {Starnini},\ and\ \citenamefont
  {Pastor-Satorras}}]{moinet2014burstiness}%
  \BibitemOpen
  \bibfield  {author} {\bibinfo {author} {\bibfnamefont {A.}~\bibnamefont
  {Moinet}}, \bibinfo {author} {\bibfnamefont {M.}~\bibnamefont {Starnini}}, \
  and\ \bibinfo {author} {\bibfnamefont {R.}~\bibnamefont {Pastor-Satorras}},\
  }\href@noop {} {\bibfield  {journal} {\bibinfo  {journal} {arXiv preprint
  arXiv:1412.0587}\ } (\bibinfo {year} {2014})}\BibitemShut {NoStop}%
\bibitem [{\citenamefont {Scalas}\ \emph {et~al.}(2006)\citenamefont {Scalas},
  \citenamefont {Kaizoji}, \citenamefont {Kirchler}, \citenamefont {Huber},\
  and\ \citenamefont {Tedeschi}}]{scalas2006durations}%
  \BibitemOpen
  \bibfield  {author} {\bibinfo {author} {\bibfnamefont {E.}~\bibnamefont
  {Scalas}}, \bibinfo {author} {\bibfnamefont {T.}~\bibnamefont {Kaizoji}},
  \bibinfo {author} {\bibfnamefont {M.}~\bibnamefont {Kirchler}}, \bibinfo
  {author} {\bibfnamefont {J.}~\bibnamefont {Huber}}, \ and\ \bibinfo {author}
  {\bibfnamefont {A.}~\bibnamefont {Tedeschi}},\ }\href@noop {} {\bibfield
  {journal} {\bibinfo  {journal} {Physica A}\ }\textbf {\bibinfo {volume}
  {366}},\ \bibinfo {pages} {463} (\bibinfo {year} {2006})}\BibitemShut
  {NoStop}%
\bibitem [{\citenamefont {Malmgren}\ \emph {et~al.}(2008)\citenamefont
  {Malmgren}, \citenamefont {Stouffer}, \citenamefont {Motter},\ and\
  \citenamefont {Amaral}}]{malmgren2008poissonian}%
  \BibitemOpen
  \bibfield  {author} {\bibinfo {author} {\bibfnamefont {R.~D.}\ \bibnamefont
  {Malmgren}}, \bibinfo {author} {\bibfnamefont {D.~B.}\ \bibnamefont
  {Stouffer}}, \bibinfo {author} {\bibfnamefont {A.~E.}\ \bibnamefont
  {Motter}}, \ and\ \bibinfo {author} {\bibfnamefont {L.~A.}\ \bibnamefont
  {Amaral}},\ }\href@noop {} {\bibfield  {journal} {\bibinfo  {journal} {Proc.
  Natl. Acad. Sci. USA}\ }\textbf {\bibinfo {volume} {105}},\ \bibinfo {pages}
  {18153} (\bibinfo {year} {2008})}\BibitemShut {NoStop}%
\bibitem [{\citenamefont {Lloyd}(2001)}]{lloyd2001realistic}%
  \BibitemOpen
  \bibfield  {author} {\bibinfo {author} {\bibfnamefont {A.~L.}\ \bibnamefont
  {Lloyd}},\ }\href@noop {} {\bibfield  {journal} {\bibinfo  {journal} {Theor.
  Pop. Biol.}\ }\textbf {\bibinfo {volume} {60}},\ \bibinfo {pages} {59}
  (\bibinfo {year} {2001})}\BibitemShut {NoStop}%
\bibitem [{\citenamefont {Keeling}\ and\ \citenamefont
  {Grenfell}(2002)}]{distrib}%
  \BibitemOpen
  \bibfield  {author} {\bibinfo {author} {\bibfnamefont {M.~J.}\ \bibnamefont
  {Keeling}}\ and\ \bibinfo {author} {\bibfnamefont {B.~T.}\ \bibnamefont
  {Grenfell}},\ }\href@noop {} {\bibfield  {journal} {\bibinfo  {journal}
  {Proc. R. Soc. B}\ }\textbf {\bibinfo {volume} {269}},\ \bibinfo {pages}
  {335} (\bibinfo {year} {2002})}\BibitemShut {NoStop}%
\bibitem [{\citenamefont {Nishiura}\ and\ \citenamefont
  {Eichner}(2007)}]{lognormal}%
  \BibitemOpen
  \bibfield  {author} {\bibinfo {author} {\bibfnamefont {H.}~\bibnamefont
  {Nishiura}}\ and\ \bibinfo {author} {\bibfnamefont {M.}~\bibnamefont
  {Eichner}},\ }\href@noop {} {\bibfield  {journal} {\bibinfo  {journal}
  {Epidemiol. Infect.}\ }\textbf {\bibinfo {volume} {135}},\ \bibinfo {pages}
  {1145} (\bibinfo {year} {2007})}\BibitemShut {NoStop}%
\bibitem [{\citenamefont {Eichner}\ and\ \citenamefont {Dietz}(2003)}]{gamma}%
  \BibitemOpen
  \bibfield  {author} {\bibinfo {author} {\bibfnamefont {M.}~\bibnamefont
  {Eichner}}\ and\ \bibinfo {author} {\bibfnamefont {K.}~\bibnamefont
  {Dietz}},\ }\href@noop {} {\bibfield  {journal} {\bibinfo  {journal} {Am. J.
  Epidemiol.}\ }\textbf {\bibinfo {volume} {158}},\ \bibinfo {pages} {110}
  (\bibinfo {year} {2003})}\BibitemShut {NoStop}%
\bibitem [{\citenamefont {Bailey}(1956)}]{fixed}%
  \BibitemOpen
  \bibfield  {author} {\bibinfo {author} {\bibfnamefont {N.}~\bibnamefont
  {Bailey}},\ }\href@noop {} {\bibfield  {journal} {\bibinfo  {journal}
  {Biometrika}\ }\textbf {\bibinfo {volume} {43}},\ \bibinfo {pages} {15}
  (\bibinfo {year} {1956})}\BibitemShut {NoStop}%
\bibitem [{\citenamefont {Chowell}\ and\ \citenamefont
  {Nishiura}(2014)}]{ebola}%
  \BibitemOpen
  \bibfield  {author} {\bibinfo {author} {\bibfnamefont {G.}~\bibnamefont
  {Chowell}}\ and\ \bibinfo {author} {\bibfnamefont {H.}~\bibnamefont
  {Nishiura}},\ }\href@noop {} {\bibfield  {journal} {\bibinfo  {journal} {BMC
  Medicine}\ }\textbf {\bibinfo {volume} {196}} (\bibinfo {year}
  {2014})}\BibitemShut {NoStop}%
\end{thebibliography}%

\end{document}